\documentstyle[aps,prb]{revtex}

\begin{document}
\draft
\preprint{}
\title{Magic number behavior for heat capacities of medium sized classical
Lennard-Jones clusters }
\author{D. D. Frantz}
\address{Department of Chemistry, University of Waterloo, \\
    Waterloo, Ontario N2L 3G1, Canada}
\date{\today}
\maketitle
\begin{abstract}
Monte Carlo methods were used to calculate heat capacities as functions of
temperature for classical atomic clusters of aggregate sizes $25 \leq N
\leq 60$ that were bound by pairwise Lennard-Jones potentials. The parallel
tempering method was used to overcome convergence difficulties due to
quasiergodicity in the solid-liquid phase-change regions. All of the clusters
studied had pronounced peaks in their heat capacity curves, most of which
corresponded to their solid-liquid phase-change regions. The heat capacity
peak height and location exhibited two general trends as functions of cluster
size: for $N = 25$ to 36, the peak temperature slowly increased, while the
peak height slowly decreased, disappearing by $N = 37$; for $N = 30$, a very
small secondary peak at very low temperature emerged and quickly increased in
size and temperature as $N$ increased, becoming the dominant peak by $N =
36$. Superimposed on these general trends were smaller fluctuations in the
peak heights that corresponded to ``magic number'' behavior, with local
maxima found at $N = 36, 39, 43, 46$ and 49, and the largest peak found at $N
= 55$. These magic numbers were a subset of the magic numbers found for other
cluster properties, and can be largely understood in terms of the clusters'
underlying geometries. Further insights into the melting behavior of these
clusters were obtained from quench studies and by examining rms bond length
fluctuations.
\end{abstract}

\narrowtext
\twocolumn
\section{Introduction}
The investigation of small atomic and molecular clusters remains an active
area of research, both
theoretically\cite{Hoare-Pal,BBDJ,DJB,BJB,AS,HA,Northby,VSA,DWB_Morse,DMW_PES,Magic_Cv,CS_magic,IHHK_magic}
and experimentally.\cite{ESR,PWKR_Ni,PLV_magic,SWSS_magic} Computer
simulations of clusters have revealed a wide diversity of fundamental
behavior that is highly sensitive to many factors, such as the cluster
aggregate size $N$, the potential used to model the intermolecular
interactions, and even the simulation method itself. The Lennard-Jones (LJ)
potential is the most commonly used model potential for computer simulations
of cluster properties, and it is especially useful for modeling rare gas
clusters. A key aspect of this model when applied to small and
medium sized atomic clusters is that the resulting potential surfaces have
global minima that are dominated by icosahedral-like structures, with
fivefold symmetries that are markedly different than the closest-packed
structure found in corresponding bulk systems.

A major motivation for studying clusters is the insight that can be obtained in
understanding the transition from finite to bulk behavior. Because of their
high proportion of surface atoms, many clusters have physical properties that
typically exhibit a very irregular dependence on their aggregate size, with
certain sizes standing out in particular. For example, the mass spectra of
clusters typically exhibit especially abundant sizes that often reflect
particularly stable structures, specially reactive clusters, or closed
electronic shells,\cite{CS_magic,IHHK_magic,ESR,SWSS_magic} These ``magic
number'' sizes have been of great theoretical
interest,\cite{Magic_Cv,CS_magic,IHHK_magic} and much work has been done in
relating magic number sequences to cluster structure in terms of ``soft''
sphere packings, since many magic number sizes in LJ clusters such as $N = 13$,
19, 55, and 147 correspond to compact structures that are especially
stable.\cite{Hoare-Pal,Northby} Many cluster properties obtained from
simulation studies such as binding energies, root mean square (rms) bond length
fluctuations and heat capacities also show very irregular dependencies on the
cluster size that mostly correspond to their usual magic number sequences. For
example, the heat capacity peak values per atom for
LJ$_{13}$\cite{DJB,J-walker} and LJ$_{55}$\cite{Lopez,DW_LJ55,CLWB,NP} are
especially large, which can be attributed to the exceptional thermodynamic
stability of their lowest-energy isomers relative to their next higher energy
configurations. To more easily compare various cluster property size
dependencies in this report, I have generalized the concept of magic numbers
for arbitrary cluster properties to refer to those cluster sizes having
particularly enhanced values relative to their immediate neighbors. In a
previous study, I examined these generalized magic number effects in LJ clusters
ranging in size from 4 to 24 atoms.\cite{Magic_Cv} In this study, I extend the
sequence to 60 atoms, which completes the filling of the second Mackay
icosahedral shell (to $N = 55$) and just begins the filling of the third shell
(to $N = 147$).

The issue of cluster phase transitions is fundamentally important, and so has
been of great interest. Because of their small size, clusters do not have the
sharp first order transition characteristic of bulk melting, but instead show
changes occurring over a range of temperatures, typically a few degrees
K.\cite{BBDJ} Although the phase-change range generally decreases as the
cluster size increases, coalescing to the bulk transition temperature as $N$
approaches infinity, both the range and the location show much variation as a
function of $N$. Cluster solid-liquid phase-change regions can be determined
from computer simulations typically by identifying characteristic changes in
certain cluster properties. For example, cluster potential energies often
change markedly as functions of temperature in the phase-change regions,
resulting in distinct peaks in the corresponding heat capacity curves; rms
bond length fluctuations often rise abruptly with increasing temperature at
the beginning of the phase-change regions, and the fractions of quenched
cluster configurations corresponding to the lowest-energy isomers often
decrease sharply with increasing temperature in these regions. Defining the
phase-change region on the basis of such behavior is somewhat ambiguous,
though, since the temperature ranges associated with these different
properties do not always coincide, and occasionally do not even overlap. For
a given property, the phase-change region temperature range can also depend
on the simulation method; for example, for Ar$_{13}$, the rise in the rms
bond length fluctuations as calculated from canonical Monte Carlo simulations
occurred about 7 K lower than that calculated from molecular dynamics
simulations.\cite{DJB} Because of these difficulties, it is convenient to
loosely define the solid-liquid phase-change region as the temperature region
between the lower-temperature, solid-like region (where the clusters are
mostly confined to their lower energy configurations, undergoing isomerizations
only rarely at the higher-temperature end of the region) and the
higher-temperature, liquid-like region (where the clusters exhibit easy
isomerization and diffusion). Cluster behavior in the solid-liquid
phase-change regions is particularly complicated, with some cluster
sizes showing coexistence behavior, where ``solid-like'' and
``liquid-like'' isomers dynamically coexist,\cite{BBDJ,DJB,BJB} while other
cluster sizes show a relatively smooth progression of isomerizations between
locally similar but globally distinct configurations occurring throughout this
region.\cite{AS} Although clusters having magic number sizes such as 13 and 55
have pronounced heat capacity peaks that correspond closely to the solid-liquid
phase-change region determined from other cluster properties, other cluster
sizes such as $N = 15$, 16 and 17 do not have heat capacity peaks in their
phase-change regions.\cite{Magic_Cv} Since heat capcity peaks
have often been used to identify phase-change
regions,\cite{LW,Lopez,NCFD_LJ38,DWM_LJ38_Cv,DW_LJ55,CS_Na} characterizing
their temperature dependencies for sequences of LJ clusters can help elucidate
the underlying reasons for such varied behavior.

The accurate determination of cluster properties such as the heat capacity is
hampered by the poor convergence of standard simulation methods in the
phase-change regions. This poor convergence is a consequence of
quasiergodicity, or the incomplete sampling of configuration
space.\cite{Val-Whit} Various methods have been developed in recent years to
reduce the systematic errors resulting from quasiergodicity, including
histogram methods,\cite{LW} jump-walking methods
(J-walking),\cite{J-walker,JWFPI,Lopez-Freeman,Binary_Cv,Lopez,Matro_Freeman,CFD-Jwalk,XB-multican-Jwalk}
smart walking methods (S-walking),\cite{ZB-S-walking} and parallel tempering
methods.\cite{GT_PT,GC_PT,NCFD_LJ38} Many of these methods are based on the
coupling of configurations obtained from ergodic higher-temperature
simulations to the quasiergodic lower-temperature simulations. In my
previous study of magic number effects in cluster heat capacities, I used
J-walking to successfully overcome quasiergodicity. Monte Carlo J-walking
methods couple the usual small-scale Metropolis moves made by a
lower-temperature random walker, with occasional large-scale jumps that move
the random walker out of the confined regions of configuration space that it
could not otherwise easily escape within the duration of the simulation.
There are two complimentary implementations of J-walking. In tandem
J-walking, higher-temperature walkers and lower-temperature walkers are run
in tandem, with the higher-temperature walkers occasionally feeding
configurations to the lower-temperature walkers. Because J-walking obeys
detailed balance only approximately, the walkers become correlated whenever a
lower-temperature walker accepts a jump, and so it is necessary to run
several extra Metropolis passes after a jump is accepted to break these
correlations, thereby decreasing the overall efficiency. The second method
breaks the correlations by running the higher-temperature walkers beforehand,
and storing representative samples of the configurations in external files,
which the lower-temperature walkers sample periodically in a random fashion.
Thus, sampling from an external distribution is more efficient, provided that
the required distributions are not too large, and this was the implementation
I used in the previous study. It was also successful for the first few
clusters considered in this study, but for clusters ranging in size from $N =
31$ to 37, which had two heat capacity peaks, the heat capacity curves
obtained with different J-walker distributions showed poor reproducibility
for the lower-temperature peaks. Since the distribution file sizes already
totalled more than 2 Gb, it seemed prudent to abandon that implementation and
switch to the tandem J-walker implementation. However, a recent study by
Neirotti, {\it et al.}\cite{NCFD_LJ38} of LJ$_{38}$, which is notoriously
difficult to simulate accurately, showed that the parallel tempering method
was remarkably successful in overcoming quasiergodicity and calculating an
accurate heat capacity curve. Parallel tempering is nearly identical to a
parallel version of tandem J-walking, except that instead of copying the
higher-temperature J-walker's configuration to the lower-temperature walker's
configuration whenever a jump is accepted, the two configurations are
exchanged. This has the advantage of maintaining detailed balance, thereby
eliminating the need to run extra Metropolis passes after accepting an
exchange, which makes the method more efficient than tandem J-walking. Thus,
I decided to use parallel tempering for all of the cluster sizes examined in
this study.

I begin in Section~\ref{Sec:method} with a brief description of the
computational methods used in this work and a summary of the calculations
undertaken. Section~\ref{Sec:results} describes the results obtained
for the cluster potential energies, heat capacities, rms bond length
fluctuations and quenched configuration distributions, as functions of
temperature and cluster size. In Section~\ref{Sec:discussion}, I summarize
the important conclusions concerning magic number behavior in LJ clusters for
$N \leq 60$.

\section{Computational Methods}       \label{Sec:method}
The clusters studied were modeled by the pairwise additive Lennard-Jones
potential,
\begin{eqnarray}
    V & = & \sum_{i < j} V_{LJ}(r_{ij}), \nonumber \\
    V_{LJ}(r_{ij}) & = & 4\epsilon\left[\left(
        \frac{\sigma}{r_{ij}}\right)^{12}
        - \left(\frac{\sigma}{r_{ij}}\right)^6 \right].
\end{eqnarray}
Because small clusters can dissociate at higher temperatures and make the
cluster definition ambiguous, it is customary to confine the clusters by a
constraining potential. For classical Monte Carlo simulations, a perfectly
reflecting spherical constraining potential of radius $R_c$, centered on the
cluster's center of mass, is most convenient, and was used in this work.

The classical internal energy and heat capacity were calculated by the
usual expressions for an $N$-atom cluster (in reduced units),
\begin{eqnarray}
    \left\langle U^* \right\rangle & = & \frac{3NT^*}{2}
        + \left\langle V^* \right\rangle, \\
    \left\langle C_V^* \right\rangle & = & \frac{3N}{2}
    + \frac{\left\langle{(V^*)^2}\right\rangle
    - \left\langle{V^*}\right\rangle^2}{(T^*)^2}, \label{Eq:Cv}
\end{eqnarray}
where $U^* = U/\epsilon$, $V^* = V/\epsilon$, $C_V^* = C_V/k_B$, and
$T^* = k_BT/\epsilon$. Since so many rare gas cluster simulations have been
done for Ar, I found it convenient to generate the temperature-dependent
curves using Ar parameters, with $\sigma = 3.405$~\AA\ and $\epsilon =
119.4$~K.

Relative rms bond length fluctuations have been calculated in many previous
cluster simulations.\cite{BBDJ,DJB,BJB,Magic_Cv,ST_Ar7,CS_Na} These were
calculated as
\begin{equation}
    \delta = \frac{2}{N(N-1)}\sum_{i < j}
        \frac{\left(\langle r^2_{ij}\rangle
        - \langle r_{ij}\rangle^2\right)^{1/2}}
        {\langle r_{ij}\rangle},
\end{equation}
where the summations are taken over the entire walk. Because of the summations
being taken over the entire walk, rms bond length fluctuations cannot be
determined for the J-walking and parallel tempering simulations, whose
configuration jumps and exchanges effectively reduce the long Metropolis
walks to a collection of very many short excursions about the last accepted
jump or exchange configuration. Thus, quasi-ergodicity remains a problem for
rms bond length calculations. Caution must also be exercised when interpreting
$\delta(T)$ curves, since they are dependent on simulation parameters such as
the walk length, as well as the ensemble that is used.\cite{DJB,Magic_Cv,CS_Na}

\subsection{J-walking}
J-walking runs were done for cluster sizes $25 \leq N \leq 38$, using
sampling from externally stored distributions as described in the preceding
study.\cite{Magic_Cv} Distributions for each temperature consisted of several
files, totaling $2.5 \times 10^6$ configurations, sampled every 100,
50, 25, or 10 passes, depending on the temperature (100 passes for higher
temperatures having broad distributions, 10 passes for very low temperatures
having very narrow distributions). The distributions were generated in stages
from higher temperatures to lower, beginning with temperatures well within
the cluster liquid region (50 to 60 K). Data accumulation was done with walks
consisting of $2 \times 10^6$ total passes, with jumps attempted randomly
with a frequency of 10\%. Since any systematic errors in the higher
temperature J-walker distributions can feed into the lower temperature
distributions and corrupt them, heat capacity peaks occurring at very low
temperatures can be very difficult to calculate accurately. In previous
studies of 13-atom binary clusters,\cite{Binary_Cv} very low temperature heat
capacity peaks corresponding to mixing anomalies could be obtained accurately
by repeating the full J-walking temperature scan several times, each time
with newly generated distributions, and averaging the independent curves.
This procedure was followed in this work as well, and at least five J-walking
runs were done for cluster sizes 25 to 30, with the heat capacity curves
showing very good reproducibility in each case. For cluster sizes 31 to 38
though, the curves showed very poor reproducibility for the lower-temperature
peaks, and so the method was abandoned.

Despite the J-walking method's inability to reproduce the lower-temperature
heat capacity peaks accurately, the method was still useful for obtaining a
listing of the important lowest-energy configurations for each cluster size.
In each case, representative subsets of the distribution files were quenched
using steepest descent methods\cite{SW} and all of the unique isomers found were
extracted. While this crude method cannot identify all of the configurations
for clusters of these sizes, it is well suited for identifying most of the
lower energy isomers that dominate the low temperature behavior of clusters.
For example, the quenched J-walker distributions for LJ$_{38}$ contained all of
the lowest-energy configurations found by Doye {\it et al.},\cite{DMW_LJ38}
including the global minimum face-centered-cubic (fcc) truncated octahedron,
which is notoriously difficult to obtain from randomly initialized simulations.
These findings imply that J-walking was successful in accessing all of the
important regions of configuration space for these clusters, even though it was
unable to do so in the statistically representative manner needed to accurately
reproduce the heat capacities at very low temperatures. Thus, higher
temperature J-walker distributions were also generated for cluster sizes 39 to
60 to obtain sets of the important low-energy quenched configurations.

\subsection{Metropolis Monte Carlo}
Metropolis simulations were also run for each cluster. These provided a check
of the J-walking and parallel tempering results for those temperature regions
where quasiergodicity in the Metropolis runs was not a problem, as well as
revealing trends in the systematic errors arising from quasiergodicity in
the Metropolis runs themselves. Temperature scans were also generated using
Ar parameters, with the temperature mesh size set to $\Delta T = 1$~K, and
sometimes to 0.5~K in the phase-change regions. For each temperature,
simulations consisted of $10^5$ warmup passes, followed by $10^7$ passes with
data accumulation. The scans in each case were started at the lowest
temperature from the global minimum configuration obtained from the
corresponding J-walker simulations, and were continued well past the cluster
solid-liquid phase-change region, with the final configuration for each
temperature used as the initial configuration for the subsequent temperature.
To gain a better appreciation of the extent of quasiergodicity in the
solid-liquid phase-change region, additional Metropolis simulations
consisting of $10^8$ total passes per temperature point were run for each
cluster over a narrower temperature range encompassing the phase-change
region.

\subsection{Parallel Tempering}
Parallel tempering\cite{GT_PT,GC_PT,NCFD_LJ38} is very closely related to the
tandem and parallel walker implementations of
J-walking,\cite{J-walker,Matro_Freeman} with the essential difference
being that instead of the higher-temperature configuration being copied over
the lower-temperature configuration whenever a jump is accepted, the two
configurations are exchanged. This close similarity between parallel
tempering and tandem J-walking makes it very easy to convert a tandem
J-walking simulation program to a parallel tempering one. The parallel
tempering method's configuration exchange ensures that it obeys detailed
balance (provided that the simulation has been warmed up enough to be in the
asymptotic region), so there is no need to run the extra Metropolis passes to
break correlations between the higher-temperature and lower-temperature
walkers whenever an exchange move is accepted, which is required for tandem
J-walking. In addition to offering greater efficiency, the obeying of
detailed balance allows parallel tempering simulations to vary the
exchange-attempt frequency in temperature regions where the
exchange-acceptance rate is too low. In J-walking simulations, the
jump-attempt frequency was typically kept at 10\%, since larger values
required even more extra Metropolis passes to be run whenever a jump was
accepted to break the correlations, thereby offsetting the efficiency gains
made by increasing the jump-attempt frequency in the first place.

As with J-walking, the number of temperatures used in a parallel
tempering simulation and their spacing cannot be chosen arbitrarily, but must
be selected to ensure that the exchanges are accepted sufficiently often.
Generally, more temperature walkers and smaller spacing are required for the
solid-liquid phase-change region, and for very low temperatures. Since the
heat capacity curves as a functions of temperature typically had peaks in the
solid-liquid phase-change regions whose location and height I wanted to
determine accurately, the number of temperatures and their spacing was also
affected by the need to properly represent the heat capacity curves with a
sufficient density of points to allow for good interpolation. This latter
requirement resulted in my choosing temperature spacings in the phase-change
region that were small enough that the exchange acceptance ratios were
typically quite high. For very low temperatures, the classical heat capacity
curve is nearly linear, thus requiring fewer points to represent. However,
the exchange acceptance ratio decreases rapidly at lower temperatures,
requiring more temperature walkers. Since the overall computation time is
nearly proportional to the number of walkers, I was able to reduce the number
of lower-temperature walkers needed by increasing the exchange-attempt
frequency $P_X$ at lower temperatures, according to the relation
\begin{equation}
	P_X(T) = \frac{P_X(T_H)}{1 -
				\left(\frac{P_X(T_L) - P_X(T_H)}{P_X(T_L)}\right)
				\left(\frac{T_H - T}{T_H - T_L}\right)},
	\label{Eq:Exchange}
\end{equation}
where $P_X(T_L) = (1 + 3P_X(T_H))/4$. This expression increases the
lowest-temperature exchange-attempt frequency $P_X(T_L)$ by a quarter of the
range above the baseline highest-temperature exchange-attempt
frequency, $P_X(T_H)$, which was set to 10\%.

The parallel tempering simulations done in this study were similar to those
done by Neirotti {\it et al.} for LJ$_{38}$.\cite{NCFD_LJ38} For each
cluster simulation, a number of random walkers at various temperatures were
run in parallel, with each temperature walker occasionally attempting an
exchange move with the next higher temperature walker, except for the highest
temperature walker, which attempted no exchanges directly (but did exchange
configurations according to attempts made by the next-highest temperature
walker). Since I had the lowest-energy isomers for each cluster from the
J-walking distribution quenches, for most of the clusters I initialized
the temperature walkers by using the lowest-energy isomer for the
lowest-temperature initial configuration, the next lowest-energy isomer for
the next lowest-temperature initial configuration, and so on. For the other
clusters, I just used the lowest-energy isomer to initialize all of the
initial configurations. It did not seem to matter which initialization scheme
was used, and of course, random configurations could have been used, but at
the cost of longer warmup periods. Because I initialized the simulations from
the lowest-energy isomers, I could quickly run preliminary simulations that
gave qualitatively good results, which I could use to fine-tune the number of
parallel tempering temperature walkers and their spacing. These preliminary
parallel tempering runs consisted of $10^5$ Metropolis warmup passes for each
temperature walker, followed by $10^5$ parallel tempering warmup passes and
then by 10 parallel tempering walks with preliminary data accumulation, with
each walk being $10^5$ passes in length. The configurations for each
temperature walker were saved at the end of the warmup period, and at the end
of each walk. The heat capacity curves for each walk were examined for
consistency, and if significant variations were found, the parallel tempering
run was repeated (without the Metropolis warmup passes) from the last saved
configurations, so that in effect, all of the preceding parallel tempering
runs served as warmup runs for the final run. For some of the clusters, only
parallel tempering warmup passes were run, indicating that the preceding
Metropolis warmup passes were likely not necessary (although their
computational expense was minor). After the preliminary runs appeared to be
sufficiently warmed up, a longer parallel tempering run of $10^5$ warmup
passes followed by $10^7$ total passes of data accumulation per temperature
point was run (10 walks of $10^6$ passes each). Again, the heat capacity
curves for each walk were checked for consistency; for a few clusters, the
variations in the curves were still too large, and so the run was repeated
from the last saved configurations. For LJ$_{38}$, an especially long warmup
was required, consistent with the findings of Neirotti {\it et
al}.\cite{NCFD_LJ38}

One important difference between my implementation of parallel tempering and
that of Neirotti {\it et al} concerns the choice of the constraining radius
$R_c$. Ideally, the value of the constraining radius should be large enough
that the cluster properties being simulated are not adversely affected by
the constraining potential, but not so large that cluster dissociations lead
to overly long convergence times. In my previous study of clusters of size 4
to 24, I used $R_c = 4.0\,\sigma$, a value large enough not to affect the
heat capacity curves, except at higher temperatures, well past the
solid-liquid phase-change regions. Thus, for this study, I also set $R_c =
4.0\,\sigma$ for the Metropolis and J-walker simulations. Neirotti {\it et
al} used a value of $R_c = 2.25\,\sigma$ for their simulation of LJ$_{38}$,
since they had difficulties attaining ergodicity with larger constraining
radii. For LJ$_{38}$, such a small constraining radius was not a problem
since the heat capacity peak occurred at a relatively low temperature.
However, for most of the cluster sizes included in this study, such a small
value would affect the heat capacity peak height and location. After some
preliminary investigations, I chose to set $R_c = 3.5\,\sigma$ for parallel
tempering simulations of cluster sizes 25 to 48, and to $R_c = 4.0\,\sigma$
for the larger cluster sizes. Figure~\ref{Fig:Ar30_Rc} illustrates the
effects of the constraining radius on clusters Ar$_{30}$ and Ar$_{32}$. The
heat capacity peak for $R_c = 3.0\,\sigma$ is slightly truncated, but the
peak for $R_c = 3.5\,\sigma$ is in good agreement with the $R_c =
4.0\,\sigma$ peak, even though the high-temperature side of the peak is
substantially affected. These two examples also show how well parallel
tempering was able to overcome the quasiergodicity evident in the Metropolis
curves. The two parallel tempering curves for Ar$_{32}$ show very good
agreement for the smaller peak, which is completely absent in the Metropolis
simulations.

The converged heat capacity curve I obtained for Ar$_{38}$ was mostly similar
to the curve obtained by Neirotti {\it et al.} in their parallel tempering
study of LJ$_{38}$,\cite{NCFD_LJ38} with a small shoulder in the
low-temperature side of the peak. The reduced peak temperature 0.1653 was also
in good agreement with that found by Neirotti {\it et al.} (0.166), although my
peak value was slightly higher ($\langle C_V\rangle/N k_B = 5.858$ compared to
5.62) and the higher-temperature part of my curve rose substantially higher;
these discrepancies are mostly due to Neirotti {\it et al.} having used the
smaller constraining radius. As pointed out by Neirotti {\it et al.}, the
shoulder in the heat capacity curve is in contrast to the small
lower-temperature peak found by Doyle, Wales and Miller in their study of
LJ$_{38}$\cite{DWM_LJ38_Cv} (their larger peak location and height were similar
to the parallel tempering results, though, at 0.17 and 5.68, respectively).

Finally, follow-up parallel tempering simulations of $10^6$ passes per
temperature were run for each cluster with periodic quenching of the
configurations to determine the population distributions of the lower-energy
isomers as functions of temperature. These were initialized using the final
configurations from their corresponding preceding parallel tempering runs.
Since I was primarily interested in using the quench results to provide
qualitative insight into the heat capacity results, rather than determining
the quench curves themselves with high accuracy, these shorter walk lengths
were sufficient.

\section{Results}       \label{Sec:results}
\subsection{Structural properties}
Much of the magic number behavior of cluster heat capacities can be
understood in terms of the magic number behavior of the underlying potential
surfaces. For example, for the small clusters I studied
earlier,\cite{Magic_Cv} the largest heat capacity peaks were observed for
sizes $N = 13$, 19 and 23, corresponding to single, double and triple
icosahedra, respectively. The same magic numbers were found in some
corresponding structural properties as well, such as the binding energy
differences,
\begin{equation}
    \Delta E_b(N) = -[V_{\rm min}(N) - V_{\rm min}(N-1)]. \label{Vmin}
\end{equation}
The lowest-energy configurations for each of the clusters I examined in this
study were obtained from quenches of J-walker distributions. These were
consistent with the lowest-energy configurations found in other
studies.\cite{Northby,DWM_LJ38_Cv} From $N = 25$ to 30, the global minima all
consisted of a $N = 19$ double icosahedral core with atoms added to
tetrahedrally bonded face sites (anti-Mackay overlayers), while for $N = 31$ to
54 (except for $N = 38$), the added atoms occupied sites corresponding to the
outer shell of the $N = 55$ Mackay icosahedron (Mackay
overlayers);\cite{Northby,DWB_Morse} $N = 38$ did not fit the pattern, it
being a truncated octahedron instead.\cite{NCFD_LJ38,DWM_LJ38_Cv} Clusters for
$N = 56$ to 60 had added atoms that contributed to the anti-Mackay overlayer of
the 55-atom core, beginning the building up of the next Mackay icosahedron ($N
= 147$).

Although the lowest-energy isomers play a significant role in cluster
thermodynamics, other low-energy isomers also play important roles, and the
size of the gap between the lowest-energy isomer and the next lowest-energy
isomer is an especially important predictor of magic number
behavior.\cite{BBDJ,HA,DWB_Morse} Figure~\ref{Fig:EnergyLevels} shows the
energy levels of the lower-energy configurations for each cluster size,
relative to their global minimum. Sizes $N = 26, 29, 32, 36, 39, 43, 46, 49,
55$ and 58 have relatively large gaps between their global minimum and next
higher minimum, compared to their immediate neighbors, and so can be
considered a magic number sequence; the gaps between the lowest-energy and
next lowest-energy isomers are also plotted in Figure~\ref{Fig:EnergyN} as
functions of $N$. The Mackay icosahedron $N = 55$ has an especially large
gap. There is also much variation in the densities of the low-lying isomers
evident in Figure~\ref{Fig:EnergyLevels}, with densities being substantially
greater for sizes $N = 28$ to 37 than for the other sizes, while sizes $N =
52$ to 56 have particularly low densities.

The binding energy differences for the clusters are also displayed in
Figure~\ref{Fig:EnergyN}. Except for $N = 38$, the maxima in the binding
energy differences are the same as the maxima in the energy differences
between the two lowest-energy isomers. Also shown in Figure~\ref{Fig:EnergyN}
is a plot of another indicator of structural magic number behavior, the
second finite difference of the energy,
\begin{equation}
	\Delta_2 E(N) = V_{\rm min}(N + 1) + V_{\rm min}(N - 1) - 2
						V_{\rm min}(N).
\end{equation}
These magic numbers are also in agreement with the magic numbers obtained for
the binding energy differences, again, except for $N = 38$. The discrepancy
in the magic number sequence for $N = 38$ is due to that cluster being unique
in this range by having an fcc structure as its global minimum, so that those
measures that are relative to the cluster's immediate neighbors are
considerably different than those of the other clusters, whose immediate
neighbors are structurally similar.

Figure~\ref{Fig:Del2E_NT} shows a generalization of the zero-temperature
$\Delta_2 E(N)$ curve shown in Figure~\ref{Fig:EnergyN}, with the curves
extended to non-zero temperatures to form a $\Delta_2 E(N,T)$ surface. The
curves were obtained from parallel tempering simulations; the temperature scale
is for Ar. The surface can be seen to encompass two general regions: a flat
region at higher temperatures, corresponding to the clusters' liquid-like
regions, and a corrugated region at lower temperatures, corresponding to the
clusters' solid-like regions. The plot shows that many of the clusters are
locked in their lowest-energy isomers for extended temperature ranges,
especially the magic number sizes $N = 26$, 29, 39, 43, 46, 49 and 55 --- the
$\Delta_2 E$ peak values differ very little from their zero-temperature values
well into the phase-change region, until their abatement over a relatively
narrow temperature range. The plot also shows that the phase-change temperature
regions have a strong dependence on the cluster size, with the melting
temperature generally decreasing from about 30 K for $N = 26$ to about 20 K for
$N = 36$, and then generally increasing to a maximum of about 40 K for $N =
55$.

\subsection{Thermodynamic properties}
Figure~\ref{Fig:potential} plots the reduced potential energy per particle
for each cluster size over a temperature range spanning the solid-like and
liquid-like regions. Magic number behavior can be seen for sizes $N = 26$,
29, 32, 36, 39, 43, 46, 49 and 55 by the relatively large differences at
lower temperatures between their potential energies and those of their
preceding cluster size. The Mackay icosahedron $N = 55$ is so stable
relative to its neighbors that its curve actually crosses several of the
other curves. The phase regions are apparent in Figure~\ref{Fig:potential} by
the transition from the irregular spacing between the curves seen for the lower
temperature, solid-like regions, to the regular spacing seen for the higher
temperature, liquid-like regions. As was seen in the $\Delta_2 E(N,T)$ surface
in Figure~\ref{Fig:Del2E_NT}, the melting temperature generally decreases as
the cluster size increases from $N = 25$, reaching a minimum of about 20 K for
$N$ about 36, and then increases to a maximum for $N = 55$.

Curves of the reduced heat capacity per particle for each cluster size are
shown in Figures~\ref{Fig:Cv1}, \ref{Fig:Cv2} and \ref{Fig:Cv3}, while
Figure~\ref{Fig:magic} shows the corresponding peak height and temperature as
functions of the cluster size; the peak parameters are also listed in
Table~\ref{Tbl:Heat Capacity Peak}. The peak parameters were obtained by
smoothing and interpolating the combined parallel tempering data (10 walks of
$10^6$ passes per temperature point each). The associated parameter
uncertainties were estimated by determining the peak height and location for
each of the 10 walks separately, and calculating the standard error of the
set of these values. As a check, additional runs of $10^7$ total passes per
temperature were done for a few of the clusters, and the standard deviations
of the resulting set of peak parameters were compared to the corresponding
uncertainty estimates. These were found to be in good agreement. The extra
check runs also showed the parallel tempering heat capacity results to be
very reproducible, unlike the corresponding J-walking results.

Each of the clusters had at least one peak in its heat capacity curve, and
overall, while their was much variation in the curves, systematic trends are
evident. Cluster sizes $N = 36, 39, 43, 46, 49$ and 55 had peak heights that
were substantially greater than those of their immediate neighbors, and thus
represent the magic numbers for the heat capacity sequence. These cluster
sizes were also magic numbers for the $\Delta_2 E(N)$ sequence and the
lowest-energy isomer gaps shown in Figure~\ref{Fig:EnergyN}, but the other
structural magic numbers ($N = 26, 29$ and 32) did not show magic number
behavior in their heat capacity curves. Instead, for cluster sizes 25 to 36,
both the heat capacity peak height and the peak temperature showed remarkably
smooth variation, with the peak temperature slowly increasing, while the peak
height slowly decreased in a nearly linear fashion, until the peak
disappeared at $N = 37$. As can be seen in Figure~\ref{Fig:Cv1}, this
behavior was somewhat affected by the choice of the constraining radius. For
the Metropolis simulations run at $R_c = 4.0\,\sigma$, the heat capacity
curves had greater slopes at higher temperatures, resulting in the peak
disappearing by $N = 34$.

More interesting was the emergence and evolution of a second heat capacity
peak, as the first peak gradually disappeared. Beginning with $N = 30$, a
small, lower-temperature peak can be seen in Figure~\ref{Fig:Cv1} to develop
and gradually increase in both height and temperature, until it becomes the
dominant peak at $N = 36$. This trend can be seen to continue in
Figures~\ref{Fig:Cv2} and \ref{Fig:Cv3}, until the peak reaches its greatest
height and highest temperature for the magic number size $N = 55$. This
behavior suggests partitioning the heat capacity peaks into two sequences,
with the first sequence containing the small-$N$ peaks that evolve into the
large, broad high-temperature peaks as the second Mackay shell initially
builds up, while the second sequence originates from the smaller, narrower,
lower-temperature peaks that evolve into the dominant peaks as the second
Mackay shell is completed. The possible beginning of a third sequence
corresponding to the initial building up of the third Mackay shell can be
inferred from the curves in Figure~\ref{Fig:Cv3}, where Ar$_{58}$ has a
small, lower-temperature peak, and Ar$_{59}$ and Ar$_{60}$ have discernable
shoulders in the low-temperature sides of their peaks.

The emergence of the smaller peak nearly coincides with the transition at
$N = 31$ from the anti-Mackay overlayer in the lowest-energy configuration to
the Mackay overlayer, and the cluster sizes where two prominent heat capacity
peaks are evident correspond to those sizes having the greatest density of
low-lying isomers (Figure~\ref{Fig:EnergyLevels}). Doye {\it et al.} found that
for clusters in this size range, the high density of isomers was due to the
small differences in energy between the various configuration types,
anti-Mackay, Mackay, decahedral, fcc, etc;\cite{DMW_PES} similar findings were
obtained for clusters bound by a Morse potential whose range parameter is
similar to the Lennard-Jones potential.\cite{DWB_Morse}

The two heat capacity sequences make the determination of the cluster melting
temperature from heat capacity peak temperatures ambiguous for sizes $N = 31$
to 36. Peaks in heat capacity curves have typically been associated with
cluster phase changes, particularly solid-liquid
changes.\cite{LW,Lopez,NCFD_LJ38,DWM_LJ38_Cv,DW_LJ55,CS_Na} However, peaks can
arise from other circumstances as well, such as mixing
anomalies\cite{Lopez-Freeman,Binary_Cv} and premelting effects, which can make
the assignment of the cluster melting temperatures difficult. Calvo and
Spiegelmann defined the melting temperatures of the sodium clusters they
studied whose heat capacity curves had multiple peaks, as the temperature
associated with the highest peak.\cite{CS_Na} Applying such a definition to the
clusters shown in Figure~\ref{Fig:Cv1} that have two heat capacity peaks would
imply that the reduced melting temperature would jump from $T^* = 0.3784$ for
$N = 35$ to 0.1427 for $N = 36$. Given the structural similarities of these two
clusters, such a large difference in the melting temperature would be
surprising, and as will be seen later when other cluster properties are
examined, the two clusters do in fact have similar melting temperatures. Also,
the potential energy curves shown in Figure~\ref{Fig:potential} and the
$\Delta_2 E(N,T)$ surface shown in Figure~\ref{Fig:Del2E_NT} both suggest that
the melting temperature for these clusters is about $T^* = 0.17$, and thus
intermediate to the two heat capacity peak temperatures.

There is much evidence of systematic error due to quasiergodicity in the
Metropolis results, especially for sizes $N = 31$ to 38, where the Metropolis
heat capacity curves had lower-temperature peaks that were either missing ($N
= 31$ to 34) or poorly formed ($N = 35$ to 38). For the other cluster sizes,
the Metropolis heat capacity curves showed very good agreement with the
parallel tempering results, except for temperatures near the heat capacity
peak, where the Metropolis results for $10^7$ total passes per temperature
point were substantially lower than the parallel tempering results, and even
the Metropolis results for $10^8$ total passes were typically slightly lower.
The fact that parallel tempering simulations of $10^7$ passes could provide
significantly better results than similar Metropolis simulations of $10^8$
passes, for not much more computational overhead than Metropolis simulations
of $10^7$ passes, clearly shows the superiority of parallel tempering over the
Metropolis method.

There have been several previous studies of 55-atom cluster heat~capacities
done.\cite{LW,Lopez,DWM_LJ38_Cv,DW_LJ55,CLWB,NP} The study most directly
comparable to the parallel tempering results I obtained for $N = 55$ is a
J-walking study of Ar$_{55}$ done by L\'{o}pez.\cite{Lopez} He obtained peak
parameters of $\langle C_V \rangle/N k_B = 19.5$ and $T^* = 0.29$, compared
to my parallel tempering results of 17.45 and 0.294, respectively. L\'{o}pez
used the same constraining volume radius as I did for $N = 55$, $R_c =
4.0\,\sigma$, but his J-walker distributions consisted of only $4 \times
10^4$ configurations, which are considerably fewer than the $2.5 \times 10^6$
configurations I used for my J-walking studies of cluster sizes $N = 25$ to
38. Given the poor results I obtained with J-walking using larger
distributions for smaller cluster sizes, I suspect that the discrepancy in
the heat capacity peak parameters is due to L\'{o}pez having used
insufficiently representative J-walker distributions in his study. The heat
capacity curve obtained from a Metropolis simulation of $10^8$ total passes
per temperature that is shown in Figure~\ref{Fig:Cv3} had peak parameters of
17.00 and 0.294, respectively, which are close to the parallel tempering
results. The small discrepancy between these Metropolis results and the
parallel tempering results is similar to the discrepancies obtained for the
other clusters of similar size. Labastie and Whetten obtained peak parameters
of 15.4 and 0.3 for Ar$_{55}$ using histogram methods, but their smaller peak
height was most likely due to their constraining radius being considerably
smaller at $R_c = 2.6\,\sigma$. Doye and Wales obtained peak parameters of
17.4 and 0.298 for Ar$_{55}$ using the superposition
method,\cite{DWM_LJ38_Cv,DW_LJ55} which is in very good agreement with my
results.

\subsection{Bond length fluctuations}
The rms bond length fluctuation $\delta$ is another measure that has been used
often in cluster simulations to estimate the melting
temperature.\cite{BBDJ,DJB,BJB,Magic_Cv,ST_Ar7,CS_Na} According to the
Lindemann criterion, melting occurs in bulk matter when $\delta$ is about 10\%
or greater. However, much care must be exercised in interpreting rms bond
length fluctuations obtained from cluster simulations. First, as Calvo and
Spiegelmann have pointed out, the Lindemann criterion was originally proposed
for bulk systems, implying that it does not necessarily extend to finite
clusters, and the choice of  10\% for the critical value is somewhat
arbitrary.\cite{CS_Na} Second, $\delta(T)$ is dependent on the simulation walk
length, with the sharp rise in the curve generally occurring at lower
temperatures as the walk length is increased.\cite{Magic_Cv} For cluster
simulations of very long length, a value of $\delta \approx 0.1$ can be
obtained at temperatures where the cluster spends most of its time in some
solid-like form, with only occasional isomerizations to other solid-like forms,
rather than undergoing the ready diffusion typical of liquid-like behavior.
Because $\delta$ is obtained from averages over entire Metropolis walks or
molecular dynamics trajectories, it cannot be estimated from simulations where
configuration scrambling occurs, such as J-walking and parallel tempering, and
so quasiergodicity issues further complicate matters. Using very long
walk-lengths to reduce systematic errors arising from quasiergodicity can
result in $\delta(T)$ curves whose transitions occur well before the melting
phase-change.

Despite these caveats, rms bond length fluctuation curves have qualitative
validity and can provide useful insights, especially in conjunction with
other cluster properties. Curves of $\delta(T)$ for each cluster size are
shown in Figures~\ref{Fig:rms1}, \ref{Fig:rms2} and \ref{Fig:rms3} for
averaged walk-lengths of $10^6$ and $10^7$ passes per temperature, while
Figure~\ref{Fig:rmsN} shows the $N$-dependence of the temperatures
corresponding to the Lindemann-like threshold $\delta = 0.2$. The threshold
temperatures were estimated by smoothing and interpolating the $\delta(T)$
curves, and then searching for the interpolated temperature corresponding to
$\delta = 0.2$. I chose to set the threshold slightly higher than the usual
Lindemann criterion of $\delta = 0.1$, since some of the $\delta(T)$ curves
had too much noise near $\delta = 0.1$ to obtain a reliable estimate of the
temperature corresponding to that threshold, and because the threshold is
somewhat arbitrary anyway; the value $\delta = 0.2$ was also typically near
the inflection point of the smoothed curves' transition regions.

The $\delta(T)$ curves shown in Figures~\ref{Fig:rms1} to \ref{Fig:rms3} are
mostly similar, but there are some interesting differences. Generally, the
steepness of the rise in $\delta$ in the transition region increases with
cluster size, with the magic number sizes $N = 43, 46$ and 55 having
especially sharp rises. Evidence of quasiergodicity can be clearly seen in
some curves: the $\delta(T)$ curves for $N = 31$ each have a noisy shoulder
just before the transition region, while the curves for $N = 32$ each have a
small peak; the curves for $N = 38$ show the cluster to be locked in its
global minimum for all of the lower-temperature Metropolis simulations of
length $10^7$ passes or fewer, until finally escaping at $T = 19$ K. Evidence
of pre-melting in the anti-Mackay overlayer of the third Mackay shell can be
seen in sizes $N = 56$ to 60 by the emergence and evolution of a shoulder in
the lower-temperature part of the $\delta(T)$ curves.

Also indicated on the temperature axes in Figures~\ref{Fig:rms1} to
\ref{Fig:rms3} are the corresponding heat capacity peak temperatures obtained
from the parallel tempering simulations (likewise, Figures~\ref{Fig:Cv1} to
\ref{Fig:Cv3} have the corresponding temperatures associated with the
Lindemann-like threshold of $\delta = 0.2$ indicated on their axes). These
clearly show that the heat capacity peaks do not necessarily correspond to a
cluster phase change. For cluster sizes $N = 25$ to 29, the heat capacity
peak temperatures are well within the liquid-like region, according to the
$\delta(T)$ curves, while for cluster sizes $N = 30$ to 36, where two heat
capacity peaks are evident, the higher-temperature heat capacity peak is
also clearly in the liquid-like region. Moreover, these temperatures are high
enough that quasiergodicity in these regions of the $\delta(T)$ curves is
not an issue. Likewise, most of the lower-temperature heat capacity peaks are
found to be in the solid-like region of the $\delta(T)$ curves, well before
the transition region (although ergodicity issues regarding the $\delta(T)$
curves in these regions could be involved). Only for sizes $N = 35$ and $N =
36$, are the lower-temperature heat capacity peak temperatures similar to the
$\delta(T)$ transition regions. For the remaining cluster sizes shown in
Figures~\ref{Fig:rms2} and \ref{Fig:rms3}, however, the heat capacity peak
temperatures correspond quite closely to the end of the $\delta(T)$
transition regions, where the curves quickly change slope to mark the start
of the liquid-like region.

Magic number behavior in the rms bond length fluctuations is evident in
Figure~\ref{Fig:rmsN}, especially for the threshold temperature values
obtained from the longer Metropolis simulations of $10^7$ passes. The
threshold temperatures are significantly higher for the magic number sizes $N
= 26, 29, 32, 36, 39, 43, 46, 49$ and 55, relative to their immediate
neighbors. These are the same magic numbers seen in Figure~\ref{Fig:EnergyN}
for $\Delta_2 E(N)$ and for the energy gaps between the lowest-energy and
next lowest-energy isomers, except for $N = 58$. The $\delta$ data also
account for the absence of magic number behavior in the heat capacities for
sizes $N = 26, 29$ and 32. Since the large, broad, higher-temperature heat
capacity peaks seen slowly diminishing in Figures~\ref{Fig:EnergyN} and
\ref{Fig:Cv1} are located well past the cluster solid-liquid phase-change
regions, whatever magic number effects might exist for $N = 26, 29$ and 32 are
masked by the systematic evolution of the heat capacity peak itself with
increasing cluster size. Classical cluster heat capacities are inherently
a measure of the width of the potential energy distributions. Typically, for
temperatures near the cluster melting temperature, the distributions are much
wider than those at lower temperatures (which span mostly lower-energy,
solid-like forms) and those at higher temperatures (which span mostly
higher-energy, liquid-like forms), since these distributions span both types
of forms. However, the distribution width in the liquid region also typically
increases with temperature, since the number of accessible configurations
also increases. Thus, depending on the energy distribution of the
higher-energy isomer potential minima, the inverse dependence of the heat
capacity on temperature in Eq.~(\ref{Eq:Cv}) could result in a peak occurring
in the liquid-like region of the curve that would have little relation to the
cluster's solid-liquid phase change.

The threshold temperatures corresponding to $\delta = 0.2$ in
Figure~\ref{Fig:rmsN} also imply that the cluster melting temperature
generally decreases as the cluster size increases from $N = 25$, reaching a
minimum near $N = 37$, then generally increases with cluster size, reaching a
maximum for $N = 55$, and decreases thereafter; magic number effects
superimposed on these general trends account for the local variations. These
general trends are consistent with those inferred from the $\Delta_2 E(N,T)$
surface in Figure~\ref{Fig:Del2E_NT} and the potential energy curves in
Figure~\ref{Fig:potential}. Casero and Soler calculated melting temperatures
for LJ clusters of size $N = 4$ to 34 in the microcannonical ensemble,
defining the melting temperature to be the temperature where the free
energies of the solid and liquid clusters become equal.\cite{CS_magic} Their
results indicate a generally increasing trend with cluster size, with reduced
melting temperatures for sizes $N = 25$ to 34 all above $0.3$. These results
are very similar to the temperatures seen for higher-temperature heat
capacity peak temperatures in Figure~\ref{Fig:magic}, and thus are contrary
to the melting temperatures implied by the rms bond length fluctuation data
shown in Figure~\ref{Fig:rmsN}. Since the melting temperatures deduced from
the $\delta$ data are in good agreement with the melting temperatures
obtained from the heat capacity peak temperatures for cluster sizes greater
than about $N = 35$, and these temperatures are well below $0.3$ for $N <
40$, it is likely that the reduced melting temperatures for sizes $N = 25$ to
34 are also well below $0.3$.

\subsection{Quench results}
The temperature dependence on the distribution of quenched isomers is shown
in Figure~\ref{Fig:quench_magic} for those Ar clusters whose heat capacities
showed magic number behavior, and in Figures~\ref{Fig:quench1} to
\ref{Fig:quench3} for some selected non-magic number Ar clusters. Each plot
shows the percentage of quenches to each of the three lowest-energy isomers,
as well as the total percentage of quenches to the remaining higher-energy
isomers; the data for all of these plots were all obtained from parallel
tempering simulations. Much of the behavior exhibited by these curves can be
rationalized in terms of the energy spacings and densities of the
lower-energy isomers shown in Figure~\ref{Fig:EnergyLevels}.

The plots for the magic number sizes are all very similar in form, with
quenches in the lower-temperature, solid-like region represented solely by
the lowest-energy isomer, and quenches to the other isomers only appearing at
temperatures near the phase-change region, and higher. This is a consequence
of the large gap between the lowest-energy and next lowest-energy isomers
that is typical of magic number clusters. The only significant differences
between the different cluster sizes is the temperature where the fraction of
quenches to the lowest-energy isomer drops off substantially, which increases
with cluster size, and can be seen to be very similar to the heat capacity
peak temperature.

The plots for some non-magic cluster sizes shown in Figure~\ref{Fig:quench1}
for $N < 35$ are considerably different. The curves for $N = 25$ have a
similar form to the magic number curves in Figure~\ref{Fig:quench_magic}, but
both the decrease in the fraction of quenches to the lowest-energy isomer and
the increase in the fraction of quenches to the higher-energy isomers are
much more gradual. This gradual transition is consistent with the more
gradual transition for the $\delta(T)$ curve, seen in Figure~\ref{Fig:rms1}.
Also, the transition region occurs well below the heat capacity peak
temperature (but near the transition region indicated by $\delta(T)$).
Ar$_{27}$ has two very closely spaced lowest-energy isomers that are well
below its next-higher energy isomers, and most of the lower-temperature
quenches are to the second lowest-energy isomer. The temperature where the
fraction of quenches to the second lowest-energy isomer exceeds the fraction
to the lowest-energy isomer is similar to the peak temperature of the very
small lower-temperature heat capacity peak. Again, the temperature region
where the fraction of quenches to the higher-energy isomers increases
substantially corresponds well to the transition region in the $\delta(T)$
curves, and is well below the higher-temperature heat capacity peak
temperature. The quench curves for $N = 31$ and 32 are similar to the $N =
27$ curve, with the fraction of quenches to the lowest-energy isomer dropping
rapidly at very low temperatures corresponding closely to the small heat
capacity peak temperatures, and the fraction of quenches to higher-energy
isomers gradually increasing at temperatures well below the peak temperature
of the larger, higher-temperature heat capacity peaks. For Ar$_{31}$, the
density of lower-energy isomers is so great that quenches to higher-energy
isomers dominate at temperatures even below the transition region for the
$\delta(T)$ curves shown in Figure~\ref{Fig:rms1}. The lower-temperature
shoulders appearing in the $N = 31$ $\delta(T)$ curves and the behavior of
the corresponding quench curves are indicative of pre-melting and an extended
phase-change region. Ar$_{32}$ is a magic number according to such structural
properties as $\Delta E_b(N)$ and $\Delta_2 E(N)$, but not according to its
heat capacity peak parameters. Its quench curves are also different than the
typical magic-number quench curves seen in Figure~\ref{Fig:quench_magic}. The
$N = 32$ $\delta(T)$ curves show unusual peaks in the solid region, just
before their sharp rise. This region just follows the lower-temperature heat
capacity peak, and according to the quench results, almost all of the
quenches are to the second lowest-energy isomer. This indicates that the
peaks in the $\delta(T)$ curves are a result of quasiergodicity (the
Metropolis heat capacity curves in Figure~\ref{Fig:Cv1} also show small,
spurious peaks in this region).  The quench curves for $N = 33$ and 34
indicate easy access to the other low-lying isomers at low temperatures near
the small heat capacity peak temperatures.

Figure~\ref{Fig:quench2} shows quench results for some non-magic number
clusters for $N = 37$ to 48. Ar$_{37}$, like Ar$_{31}$ and Ar$_{34}$, has an
especially high density of low-lying isomers, and its quench curves also show
a substantial fraction of quenches to these isomers at temperatures
corresponding to the heat capacity peak temperature. For $N = 38$, the
temperature region where the fraction of quenches to the lowest-energy fcc
isomer drops off and the fraction of quenches to the lowest-energy
icosahedral-like isomer (the second lowest-energy isomer) increases
corresponds to the region just before the heat-capacity peak where a shoulder
can be seen in Figure~\ref{Fig:Cv2}, which is consistent with other studies of
this cluster.\cite{NCFD_LJ38} The quench curves for $N = 41, 45$ and 48 are
similar to those for $N = 37$, since these clusters all have accessible
lower-energy isomers that are only slightly higher than the lowest-energy
isomer, but the curves for $N = 42$ are similar to the curves for the magic
number clusters in Figure~\ref{Fig:quench_magic}, indicating that its low-lying
isomers are not very accessible.

The isomer energy levels for $N = 50$ are similar to those for $N = 45$ and
48, and the corresponding quench curves shown in Figure~\ref{Fig:quench3} are
likewise similar to those for $N = 45$ and 48. Ar$_{52}$ and Ar$_{53}$ have
some isomers having closely spaced potential energies only slightly above the
lowest-energy isomer, and then very few isomers until $\Delta V_{\rm min}
\approx 2.5$. This type of isomer energy distribution is similar to the
distributions seen for binary Ar-Kr clusters, where many mixed isomer
configurations had energies similar to the lowest-energy
isomer,\cite{Binary_Cv} and the quench results are similar too, with the
low-lying isomer quench curves showing plateaus throughout much of the
solid-like temperature region. However, neither Ar$_{52}$ nor Ar$_{53}$
exhibited the small, very low temperature heat capacity peaks that were
evident in the binary cluster curves. Ar$_{56}$ has two low-energy isomers
corresponding to the two unique ways a lone atom can be positioned on a
55-atom Mackay icosahedral core, separated by a large energy gap from the
other isomers corresponding to rearrangements of the 55-atom core. The $N =
56$ quench curves show substantial fractions of quenches to both isomers
throughout much of the solid-like region until the phase-change region, when
the higher-energy isomers become accessible. For $N = 57$, there are several
lower-energy isomers below $\Delta V_{\rm min} \approx 1.0$ that were
represented in the quench distributions, but as with $N = 52$ and 53, there
was no corresponding lower-temperature heat capacity peak. Both the $N = 57$
and 59 quench curves show evidence of premelting, consistent with their
corresponding $\delta(T)$ curves in Figure~\ref{Fig:rms3}.

\section{Conclusions}       \label{Sec:discussion}
Compared to the irregularity with respect to cluster size $N$ that
characterized the properties of the small clusters of size $N \leq
24$ that I studied previously,\cite{Magic_Cv} the properties for the clusters
sized $25 \leq N \leq 60$ showed more regularity and consistency. The
lowest-energy isomers for these sizes were all dominated by icosahedral
based geometries, except for $N = 38$, whose lowest-energy isomer had fcc
geometry (the lowest-energy icosahedral-like isomer for $N = 38$ was only
slightly less stable). For the icosahedral-like clusters, configurations
having anti-Mackay overlayers were the lowest-energy isomers for $N < 31$,
while configurations having Mackay overlayers were the lowest-energy isomers
for the remaining sizes up to $N = 55$, after which the filling of the third
shell began again with anti-Mackay overlayers. Correspondingly, the heat
capacity peak parameters formed two overlapping sequences as functions of
cluster size, with those clusters having anti-Mackay overlayer lowest-energy
isomers showing a gradual evolution in their heat capacity peaks to higher
temperature and smaller size, while those clusters having Mackay overlayer
lowest-energy isomers had small, lower-temperature peaks that generally
shifted to higher temperature and grew in size as $N$ increased.

For cluster sizes $25 \leq N \leq 35$, the heat capacity peak temperatures
did not correspond well to the cluster melting temperature ranges inferred
from the temperature dependence of other cluster properties, such as the rms
bond length fluctuations and the distributions of quenched isomers. For each
of the cluster sizes from $N = 36$ to 60, though, the melting temperature
ranges obtained from the three measures were in qualitatively good agreement.
The general trend in the cluster melting temperatures as a function of
the cluster size over the range studied was an initial decrease, reaching a
minimum near $N = 36$, then an increase until a maximum at $N = 55$, where
the second Mackay shell was completed, followed by another decrease as atoms
were added to the third Mackay shell. Superimposed on this general trend were
local maxima corresponding to the magic numbers $N = 26$, 29, 32, 36, 39, 43,
46, 49 and 55. The heat capacity peak heights had a magic number series that
was a subset of this sequence, with local maxima occuring at $N = 36$, 39,
43, 46, 49 and 55; for the other magic number sizes where the heat capacity
peaks did not show magic number behavior ($N = 26$, 29, and 32), the heat
capacity peaks were located well away from the solid-liquid phase-change
regions.

The magic number sequence for the cluster melting temperatures coincided very
well with the magic number sequences obtained for structural properties such
as the binding energy difference $\Delta E_b(N)$, the second finite
difference of the energy $\Delta_2 E(N)$, and the gap between the
lowest-energy and next lowest-energy isomers. The only differences were $N =
38$ being a magic number for the $\Delta E_b(N)$ sequence instead of $N =
39$, and $N = 58$, which was a magic number for all three structural
properties, but was not for either the heat capacity or the rms bond length
fluctuation sequences. This similarity in the magic number sequences is
consistent with the widely held view that the thermodynamic behavior of
clusters in the phase-change region is dominated by their potential energy
surfaces.

The parallel tempering method was found to be especially useful for
overcoming quasiergodicity and accurately calculating the heat capacities of
these clusters, particularly for the range $N = 30$ to 38, where the
Metropolis method was hard pressed to provide even qualitatively correct
results. Even the J-walking method, which had worked very well for
smaller cluster sizes, was unable to determine heat capacities reproducibly
for clusters in this range. Given also the relatively small computational
overhead for the method compared to Metropolis Monte Carlo, parallel
tempering currently stands as the method of choice for cluster
Monte Carlo simulations.

\acknowledgments
The support of the Natural Sciences and Engineering Research Council of
Canada (NSERC) for part of this research is gratefully acknowledged. I thank
Northern Digital, Inc. for generously providing me the use of their
workstations for some of the calculations reported, and the University of
Waterloo for the use of their facilities. I also thank David L. Freeman for
helpful discussions.

\newpage
\onecolumn
\widetext
\begin{table}
\caption{Cluster heat capacity peak parameters for $25 \leq N \leq 60$. These
values were obtained by smoothing and interpolating the parallel tempering data
shown in Figs.~\protect\ref{Fig:Cv1}, ~\protect\ref{Fig:Cv2} and
\protect\ref{Fig:Cv3}, which were based on the combined results of 10 walks,
each consisting of 10$^6$ passes of data accumulation per temperature. The
uncertainties were estimated as the standard errors of the set of peak
parameters that were obtained from the interpolated heat capacity curves for each
of the 10 walks.
\label{Tbl:Heat Capacity Peak}}
\squeezetable
\begin{tabular}{p{.125in}*{2}{p{0.75in}p{0.75in}}p{.125in}*{2}{p{0.75in}p{0.75in}}}
                                            &
\multicolumn{2}{c}{Lower temperature peak}  &
\multicolumn{2}{c}{Higher temperature peak} &
                                            &
\multicolumn{2}{c}{Lower temperature peak}  &
\multicolumn{2}{c}{Higher temperature peak} \\
\cline{2-3}
\cline{4-5}
\cline{7-8}
\cline{9-10}
\multicolumn{1}{c}{N}                        &
\multicolumn{1}{c}{$T^*$}                    &
\multicolumn{1}{c}{$\langle C_V^*\rangle/N$} &
\multicolumn{1}{c}{$T^*$}                    &
\multicolumn{1}{c}{$\langle C_V^*\rangle/N$} &
\multicolumn{1}{c}{N}                        &
\multicolumn{1}{c}{$T^*$}                    &
\multicolumn{1}{c}{$\langle C_V^*\rangle/N$} &
\multicolumn{1}{c}{$T^*$}                    &
\multicolumn{1}{c}{$\langle C_V^*\rangle/N$} \\
\hline
25 &        &                            & 0.3134$\pm$0.0016 &  6.425$\pm$0.011 &
43 &        &                            & 0.1970$\pm$0.0009 &  7.815$\pm$0.032 \\
26 &        &                            & 0.3116$\pm$0.0015 &  6.380$\pm$0.012 &
44 &        &                            & 0.1986$\pm$0.0009 &  7.012$\pm$0.027 \\
27 & 0.0183$\pm$0.0004 & 2.965$\pm$0.001 & 0.3157$\pm$0.0015 &  6.313$\pm$0.011 &
45 &        &                            & 0.2051$\pm$0.0009 &  7.682$\pm$0.029 \\
28 &        &                            & 0.3242$\pm$0.0016 &  6.210$\pm$0.013 &
46 &        &                            & 0.2210$\pm$0.0006 &  9.781$\pm$0.040 \\
29 &        &                            & 0.3299$\pm$0.0025 &  6.092$\pm$0.011 &
47 &        &                            & 0.2204$\pm$0.0005 &  8.681$\pm$0.033 \\
30 & 0.0554$\pm$0.0010 & 3.059$\pm$0.002 & 0.3337$\pm$0.0020 &  5.995$\pm$0.014 &
48 &        &                            & 0.2289$\pm$0.0010 &  9.067$\pm$0.041 \\
31 & 0.0279$\pm$0.0019 & 3.443$\pm$0.079 & 0.3396$\pm$0.0018 &  5.888$\pm$0.013 &
49 &        &                            & 0.2374$\pm$0.0006 & 10.257$\pm$0.089 \\
32 & 0.0603$\pm$0.0008 & 4.115$\pm$0.035 & 0.3484$\pm$0.0023 &  5.814$\pm$0.009 &
50 &        &                            & 0.2472$\pm$0.0006 &  9.702$\pm$0.047 \\
33 & 0.0757$\pm$0.0015 & 4.336$\pm$0.063 & 0.3511$\pm$0.0033 &  5.737$\pm$0.010 &
51 &        &                            & 0.2615$\pm$0.0003 & 11.531$\pm$0.045 \\
34 & 0.0612$\pm$0.0011 & 4.202$\pm$0.024 & 0.3641$\pm$0.0021 &  5.656$\pm$0.005 &
52 &        &                            & 0.2743$\pm$0.0004 & 13.460$\pm$0.061 \\
35 & 0.1070$\pm$0.0014 & 4.759$\pm$0.031 & 0.3771$\pm$0.0064 &  5.600$\pm$0.010 &
53 &        &                            & 0.2828$\pm$0.0005 & 15.197$\pm$0.064 \\
36 & 0.1415$\pm$0.0015 & 5.715$\pm$0.067 & 0.4097$\pm$0.0074 &  5.600$\pm$0.006 &
54 &        &                            & 0.2897$\pm$0.0004 & 16.766$\pm$0.104 \\
37 &        &                            & 0.1398$\pm$0.0017 &  5.027$\pm$0.023 &
55 &        &                            & 0.2940$\pm$0.0005 & 17.450$\pm$0.098 \\
38 &        &                            & 0.1653$\pm$0.0005 &  5.858$\pm$0.030 &
56 &        &                            & 0.2929$\pm$0.0007 & 15.280$\pm$0.131 \\
39 &        &                            & 0.1948$\pm$0.0009 &  7.554$\pm$0.029 &
57 &        &                            & 0.2917$\pm$0.0005 & 13.645$\pm$0.081 \\
40 &        &                            & 0.1938$\pm$0.0004 &  7.250$\pm$0.021 &
58 & 0.1844$\pm$0.0022 & 3.721$\pm$0.005 & 0.2871$\pm$0.0010 & 11.702$\pm$0.033 \\
41 &        &                            & 0.1875$\pm$0.0011 &  6.602$\pm$0.015 &
59 &        &                            & 0.2824$\pm$0.0006 & 10.934$\pm$0.053 \\
42 &        &                            & 0.1884$\pm$0.0006 &  6.828$\pm$0.025 &
60 &        &                            & 0.2786$\pm$0.0007 & 10.340$\pm$0.073 \\
\end{tabular}
\end{table}

\begin{figure}
\caption{Effects of the constraining radius $R_c$ on cluster reduced heat
capacities per particle for simulations of Ar$_{30}$ (at left) and Ar$_{32}$
(at right). The filled circles represent parallel tempering results for $R_c
= 3.0\,\sigma$, while the squares represent parallel tempering results for
$R_c = 3.5\,\sigma$; these were obtained from walks of $10^7$ total passes per
temperature point. The error bars are two standard deviations on each side.
The dotted line and the diamonds represent Metropolis results for $R_c =
4.0\,\sigma$, which were obtained from walks of $10^7$ and $10^8$ total
passes per temperature point, respectively. The curves with different
constraining radii are in very good agreement on the low-temperature side of
the peak, but show substantial deviations on the high-temperature side of the
peak, effecting the peak height and location slightly. The differences
between the $R_c = 3.5\,\sigma$ and $R_c = 4.0\,\sigma$ peak height and
location were minor in these two cases, and insignificant for most of the
other cluster sizes whose parallel tempering results were obtained using $R_c
= 3.5\,\sigma$. The smaller lower-temperature peaks seen in the parallel
tempering curves are absent in the Metropolis curves, indicating the
Metropolis simulations were subject to substantial quasiergodicity.
\label{Fig:Ar30_Rc}}
\end{figure}

\begin{figure}
\caption{Cluster local potential energy minima relative to the global
minimum, in reduced units. Magic number behavior can be seen for $N = 26, 29,
32, 36, 39, 43, 46, 49$ and 55, which have large gaps between their global
minimum energy and their next higher energy configurations, compared to their
immediate neighbors. The potential energies were obtained from J-walking and
parallel tempering quench studies.
\label{Fig:EnergyLevels}}
\end{figure}

\begin{figure}
\caption{The binding energy difference as a function of cluster aggregate
size (top plot), the second finite difference of the energy as a
function of cluster size (middle plot), and the energy gap between the
lowest-energy isomer and the next lowest-energy isomer, as a function of
cluster size (bottom plot), all in reduced units. Except for $N = 38$ and 39,
the three measures show the same peak values. The binding energy differences
and the second finite differences for $5 \leq N \leq 24$ have been included
for comparison, using data from Ref.~\protect\onlinecite{Magic_Cv}.
\label{Fig:EnergyN}}
\end{figure}

\begin{figure}
\caption{The second finite difference of the energy $\Delta_2 E$ as a function
of temperature $T$ and size $N$, for Ar clusters. The structure of the zero
temperature curve $\Delta_2 E(N)$ (also shown in
Fig.~\protect\ref{Fig:EnergyN}) can be seen to extend well into the
higher-temperature regions for the magic number sizes (seen as ridges in the
surface), which indicates the persistence of their solid-like lowest-energy
isomers over relatively large temperature ranges. The flat regions of the
surface at higher temperature correspond to the cluster liquid-like regions,
and so the phase-change regions correspond to the temperatures where the ridges
abate.
\label{Fig:Del2E_NT}}
\end{figure}

\begin{figure}
\caption{Cluster potential energies per particle as functions of temperature,
obtained from parallel tempering simulations for clusters ranging from $N =
25$ (top curve) to $N = 60$ (bottom curve). Reduced units have been used; the
absolute temperature scale at the top is for Ar. Magic number sizes for $N =
26$, 29, 32, 36, 39, 43, 46, 49 and 55 are indicated by the dotted curves,
and correspond to those clusters having relatively large energy differences
from their preceding cluster sizes at low temperatures. A rough indication
of the phase-change region from solid-like to liquid-like can be ascertained
by the change from the irregular spacing between the curves seen at lower
temperatures to the regular spacing seen at higher temperatures.
\label{Fig:potential}}
\end{figure}

\begin{figure}
\caption{Reduced heat capacities per particle as functions of temperature for
Ar clusters of size $N = 25$ to 36. The open circles and thicker solid
line represent the results obtained from parallel tempering runs of $10^7$
passes per temperature point. The dotted line represents the results obtained
from Metropolis runs of $10^7$ passes per temperature point, while the thin
solid line represents the results obtained from similar Metropolis runs of
$10^8$ passes per temperature point. Beginning with Ar$_{30}$, a smaller,
lower-temperature peak can be seen evolving in the parallel tempering curves
and moving to higher temperatures, while the higher-temperature peak
gradually disappears. Quasiergodicity is evident in the Metropolis results
of these cluster sizes by the absent and malformed lower temperature peaks.
As in Fig.~\protect\ref{Fig:Ar30_Rc}, the discrepancies between the parallel
tempering and Metropolis curves for higher temperatures are due to the
parallel tempering simulations having been run with a smaller constraining
radius of $3.5\,\sigma$, instead of the value of $R_c = 4.0\,\sigma$ used in
the Metropolis simulations. The large ticks on the temperature axes indicate
the temperatures where the Metropolis rms bond length fluctuations rose
sharply for walks of $10^7$ passes per temperature (higher temperature) and
$10^8$ passes.
\label{Fig:Cv1}}
\end{figure}

\begin{figure}
\caption{Same as Fig.~\protect\ref{Fig:Cv1}, but for clusters ranging
in size from $N = 37$ to 48. Note the different heat capacity scale.
\label{Fig:Cv2}}
\end{figure}

\begin{figure}
\caption{Same as Fig.~\protect\ref{Fig:Cv1}, but for clusters ranging
in size from $N = 49$ to 60. For these sizes, the constraining radius was set
to $4.0\,\sigma$ for both the parallel tempering and Metropolis simulations.
Again, note the different heat capacity scale.
\label{Fig:Cv3}}
\end{figure}

\begin{figure}
\caption{Magic number behavior in cluster heat capacities. The upper plot
shows the reduced heat capacity peak values as a function of the cluster size,
while the lower plot shows the corresponding peak reduced temperatures as a
function of the cluster size; the right-hand axis is representative of Ar.
The peak parameters were obtained from interpolations of the parallel
tempering data; parameter values for $N < 25$ have been included
from Ref.~\protect\onlinecite{Magic_Cv} for added context. Two general
sequences have been identified as a consequence of the clusters in the range
$30 \leq N \leq 37$ having two prominent peaks. The circles indicate the heat
capacity parameters for the broader, higher temperature peaks, which have
evolved from the solid-liquid phase-change region peaks of the smaller
cluster sizes, while the squares represent the narrower, lower temperature
peaks, which evolve into the solid-liquid phase-change region peaks of the
larger cluster sizes. The triangles represent the parameters for the very
small, broad peaks seen for sizes $N = 27$ and 58. The magic number labels
indicate the cluster sizes whose peaks are higher than those of their
immediate neighbors.
\label{Fig:magic}}
\end{figure}

\begin{figure}
\caption{Root mean square bond length fluctuations as functions of the
temperature for Ar clusters of size $N = 25$ to 36. The solid circles
represent data obtained from 10 Metropolis walks, each consisting of $10^6$
passes per temperature point, while the open circles represent data obtained
from 10 similar Metropolis walks of $10^7$ passes in length. The error bars
are single standard deviations for the 10 walks. The large ticks on the
temperature axes indicate the corresponding cluster heat capacity peak
temperatures.
\label{Fig:rms1}}
\end{figure}

\begin{figure}
\caption{Same as Fig.~\protect\ref{Fig:rms1}, but for clusters ranging
in size from $N = 37$ to 48.
\label{Fig:rms2}}
\end{figure}

\begin{figure}
\caption{Same as Fig.~\protect\ref{Fig:rms1}, but for clusters ranging
in size from $N = 49$ to 60.
\label{Fig:rms3}}
\end{figure}

\begin{figure}
\caption{Magic number behavior in cluster rms bond length fluctuations. The
curves plot the reduced temperatures associated with the Lindemann-like
threshold $\delta = 0.2$, as a function of cluster size. The squares
represent the values corresponding to $\delta(T)$ curves whose data were
obtained from Metropolis walks of $10^6$ passes per temperature
point, while the circles represent values obtained from similar Metropolis
simulations of $10^7$ passes. The right-hand axis is representative of Ar.
The magic numbers are a subset of those obtained in
Figure~\protect\ref{Fig:EnergyN} for the second finite differences in the
energy.
\label{Fig:rmsN}}
\end{figure}

\begin{figure}
\caption{Quenched isomer distribution curves for Ar clusters as functions of
temperature for the magic number sizes $N = 36, 39, 43, 46, 49$, and 55. The
data were obtained from parallel tempering simulations of $10^6$ total
passes per temperature. The filled circles represent the fraction of isomers
that quenched to the lowest-energy isomer, the open circles the second
lowest-energy isomer, and the squares the third lowest-energy isomer, while
the line represents the sum of the remaining isomers. The dotted vertical
lines in each plot indicate the corresponding heat capacity peak
temperatures.
\label{Fig:quench_magic}}
\end{figure}

\begin{figure}
\caption{Same as Figure~\protect\ref{Fig:quench_magic}, but for the non-magic
number sizes $N = 25, 27$ and 31 to 34.
\label{Fig:quench1}}
\end{figure}

\begin{figure}
\caption{Same as Figure~\protect\ref{Fig:quench_magic}, but for the non-magic
number sizes $N = 37, 38, 41, 42, 45$ and 48.
\label{Fig:quench2}}
\end{figure}

\begin{figure}
\caption{Same as Figure~\protect\ref{Fig:quench_magic}, but for the non-magic
number sizes $N = 50, 52, 53, 56, 57$ and 59.
\label{Fig:quench3}}
\end{figure}

\end{document}